\documentclass[letterpaper]{article}
\usepackage{isea}
\usepackage{MnSymbol}

\usepackage[pdftex]{graphicx}
\usepackage{times}
\usepackage{helvet}
\usepackage{courier}
\usepackage[numbers]{natbib}
\graphicspath{{./Figures/}}
\title{Holon: a cybernetic interface for bio-semiotics}
\author{Jon McCormack \and Elliott Wilson\\
SensiLab, Monash University\\
Melbourne, Australia\\
Jon.McCormack@monash.edu\\
}

\begin{document} 
\maketitle
\begin{abstract}
This paper presents an interactive artwork, ``Holon'', a collection of 130 autonomous, cybernetic
organisms that listen and make sound in collaboration with the natural environment. The work was developed for installation on water at a heritage-listed dock in Melbourne, Australia. Conceptual issues informing the work are presented, along with a detailed technical overview of the implementation. Individual holons are of three types, inspired by biological models of animal communication: composer/generators, collector/critics and disruptors. Collectively, Holon integrates and occupies elements of the acoustic spectrum in collaboration with human and non-human agents.

\end{abstract}


Cybernetics, Sound Art, Collective Behaviour, Emergence

\section{Introduction}

Recent advances in human-machine interfaces have shifted from direct, screen-based interaction to multi-modal dialogues, featuring new modalities, such as sound and voice. Moreover, the site of interactions is no longer confined to the office desk, home or smartphone screen: interactions take place across whole environments, sometimes between large numbers of agents, both human and non-human.

These advances are underpinned by major developments in technologies such as voice recognition, speech synthesis, digital signal processing (DSP),
machine learning (ML) and artificial intelligence (AI). However, the majority of developments in sound interaction in mainstream technologies are, understandably, linguistic -- 
used to give instructions or commands as seen in popular tools such as Google Home, Amazon Alexa and Apple's Siri. Even in more creative or experimental interactions, language often plays a dominant role \cite{Desjardins2021}. In addition, sound interactions are directed almost exclusively
to communicate \emph{to} a human audience. Speech interaction has -- understandably -- received much critical attention and criticism \cite{Porcheron2018, Reddy2021, Strengers2021}. While sound has been used to interact with the non-human, the major uses are for monitoring and identification of specific animal species (typically birds) \cite{Sethi2018, kahl2021birdnet, Vasconcelos2022}, investigation of audio enrichment for captive animals \cite{Pons2016} or audio preferences of higher primates \cite{Piitulainen2020}. Little research has been undertaken in the generation of
more general sound-based interactions between humans, other species and computers, particularly in native environments or ecological settings \cite{Brunswik1956}.

\begin{figure*}[h]
  \centering
  \includegraphics[width=\linewidth]{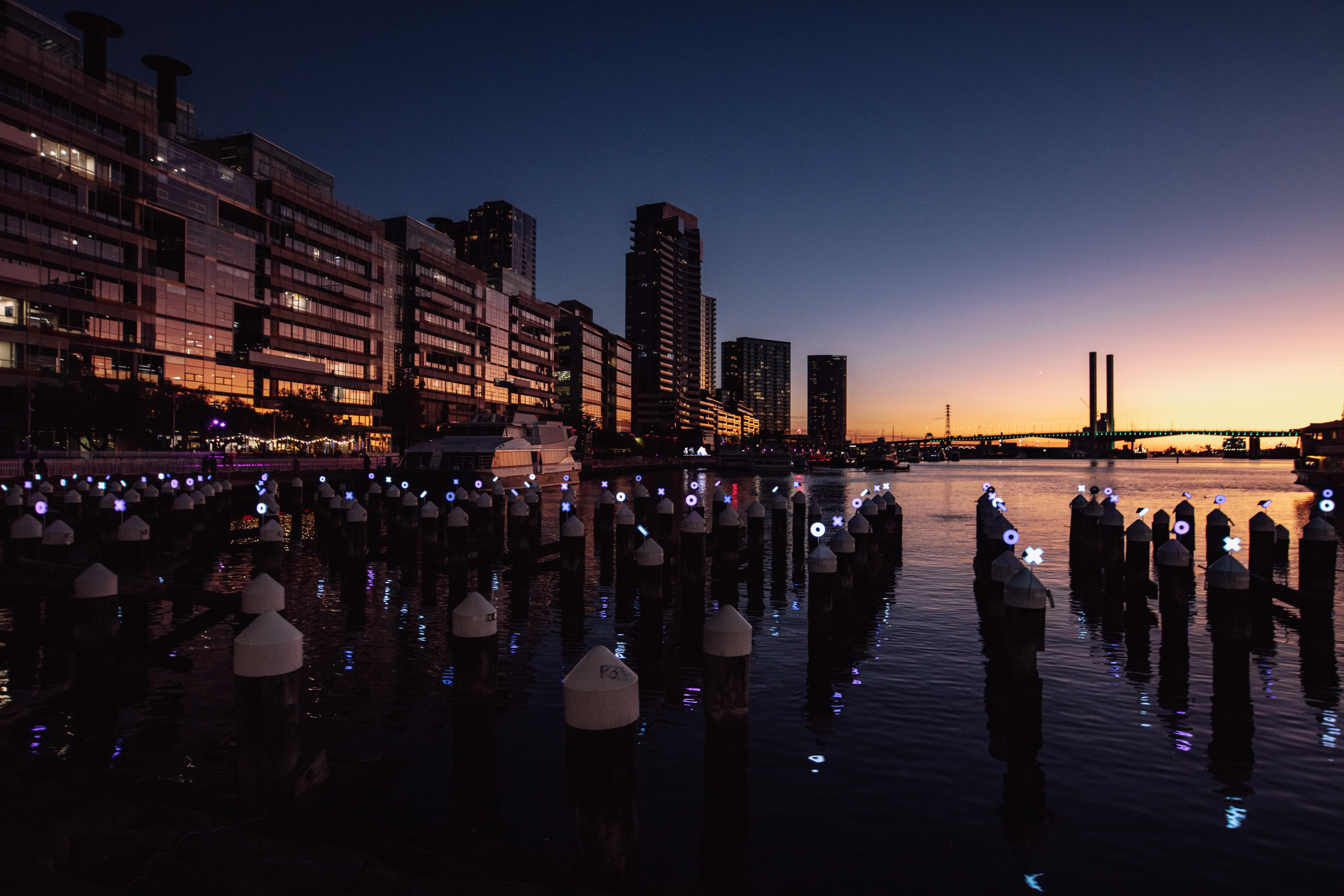}
  \caption{Holon installation at Docklands waterfront in Melbourne, August 2023. Individual holons are attached to the heritage-listed pillars (formally used in the previous century to dock ships) using a thin steel pole strapped to the white cones on top of the pillars, making them
  appear to float above the pillars. }
  \label{fig:holon}
\end{figure*}

This paper presents an art installation, \emph{Holon} (Figure \ref{fig:holon}), that takes a different approach to sound interaction. Rather than using language for commands or sound for feedback, entertainment or for information,
the work focuses on the poetic and creative possibilities for sound as a conduit  between machines and organic life, including both the human and non-human. Conceptually, the work draws upon place, environment and ecological traditions; 
rituals that bring meaning and bring a sense of community to human life allowing us to furnish time like we would furnish our home \cite{Han2020}. It brings consideration of the non-human as playing an important role in our sonic environments and soundscape ecology -- something that is increasingly being displaced or concealed by technology.

The work is informed by recent philosophical, scientific and critical studies, such as New Materialism \cite{Dolphijn2012} that de-centre the human, recognising the agency of the non-human, including other forms of life and intelligent agents in technological design \cite{Wakkary2021, Nicenboim2020} and the creative arts \cite{Aloi2021}.

\subsection{Paper Contributions}
\label{ss:contributions}
This paper presents a creative project where sound plays a dominant role as interface between technological object, humans and the non-human. ``Holon'' is a public art installation that consists of 130 individual autonomous cybernetic devices which use sound (their own mutable ``voice'') as the primary form of communication and interaction. The devices intervene on the local soundscape ecology by generating, collecting and disrupting human, animal and machine sounds.

The work highlights the experimental use of sound as the interface between different agencies in an ecological setting, highlighting how new kinds of sonic interactions can play an important role more generally as an interface beyond language-based or iconic interactions.

The work introduces the concept of ``sonic agency'' that de-centres the human to support sound exchanges between technology, humans and other biological species. The aim is to better appreciate of the role of environment and the agency of the non-human in the design of sound-based interactions.

The work's foundations arise from speculative approaches to technological design that question the formalisation of purpose or function for technological artefacts \cite{Dunne1999}. The work implicitly questions the assumptions that have dominated ``human-centred design'' for decades \cite{Norman1986} and the narratives of hedonic ``progress'' in Western culture \cite{Gray2002}. What if technology was designed not to fulfil only human desires, to not have a specific function, or solve a particular problem, to not have a defined purpose? Technological artefacts could just ``be'' in the world, as humans and other biological species are in the world \cite{Dreyfus1991, Coyne1995, Coyne1999}. Relaxing the constraints of designing for a fixed human purpose allows a more open exploration of the roles and possibilities for technological interventions in the world.

A number of philosophical and critical movements in human-machine interaction have looked at the nature of technology as a transformative mediator of human experiences and practices, such as \emph{postphemonology} \cite{Ihde1993,Verbeek2005,Verbeek2015,Hauser2018} and posthumanism \citep{Wakkary2021,Forlano2017}.
Verbeek \citep{Verbeek2015} suggests that designers ``do not merely design products, but human practices and experiences'' and therefore, ``designing things is designing human existence''. The additional step we take in the work described in this paper is to acknowledge that we are not just designing human existence, but the existence (or extinction!) of other life on Earth (either directly or indirectly) and, more speculatively, the existence of electronic or cybernetic artificial life \cite{Bedau2000}.

After examining related research and background information regarding sound as an interface in the next section, the artwork and its operation are presented in detail.

\section{Background and Related Work}
\label{s:background}

Human hearing is a highly immersive and important sense, often considered the second most important after vision \cite{schenkman2010human}. Sound has long been used in human-computer interaction for feedback. Audio icons (often referred to as \textit{Earcons}~\cite{sumikawa1985guidelines}) are brief and distinctive sounds used to represent \textit{something} \cite{Gaver1986}. They can be found, e.g.~in microwaves to announce the finish of cooking, in cars to indicate distance, in smartphones to announce the arrival of chat messages, etc. However, similar to visual icons, their communication is one-way (from device or event to ear). 

Beyond Earcons and audio feedback, audio is often used to convey more complex information, a process known as sonification \cite{Kramer1993auditory, Hermann2011sonificationhandbook, Barrett2016interactive}. This involves mapping data to different sound properties, such as pitch, volume and timbre, which can then be interpreted by the listener. But as as Ludovico et al. point out, ``many symbolic aspects of sound are culture-dependent, so it is difficult to create a sonification having a universally accepted meaning.'' \cite{Ludovico2016sonification}.

\subsection{Soundscapes}
Recently, Johansen et al. undertook an extensive review of soundscapes in interaction research, examining over 400 articles from human-computer interaction (HCI) and related literature \cite{Johansen2022}. They found two distinct categorisations of soundscape research: ``acoustic environments'' and ``compositions''. Acoustic environments are characterised by a mediation of the user's existing sound environment via perceptual construction, whereas compositional approaches used a more structured collection of sounds with direct mapping between user interactions and the composed sound. Hybridisation of these two categories was also found in the literature. Across all work -- and similar to the work presented here -- they found that the soundscape design often inherently embodied, situated and participatory. Soundscapes can also play an important role in facilitating cultural understanding, social  or environmental awareness.  For example, Andrea Polli demonstrated the effectiveness of ``soundwalks'' in promoting environmental and social awareness, supporting more nuanced social and cultural practices \cite{Polli2012}.

\subsection{Sound and Voice in Interactive Art}
Technological work in the creative arts has explored a variety of unconventional forms of sound interaction. Here we highlight projects related specifically to interaction using sound and non-linguistic modes of sound interaction.
Anne Despond and collaborators used sensors that collected atmospheric data to generate continuous music sequences via a human-composer derived mapping. These ``cloud harps'' used cloud and other atmospheric data as composition material, making them more akin to sound sculptures that use data sonification \cite{Despond2021}.

There is a long tradition from cybernetic and artificial life art that makes use of interactive sound ``agents'' -- a methodology adopted by the projects described in this paper. Inspired by concepts of emergent behaviour and bottom-up intelligence, the works use collections of sound-generating or sound-transforming agents that work collectively through local communication. Simon Penny's ``Sympathetic Sentience'' and ``Sympathetic Sentience 2'' were cybernetic devices built in the mid-1990s that received and transmitted data via sound and Infrared (IR) light \cite{Penny1999b}. Using cybernetic principles of self-governing behaviour \cite{Ashby1962} and rein control \cite{Clynes1969, Harvey2004}, the overall rhythmic build-up and flow of sound was governed through the collective emergent behaviour of 12 simple individual devices \cite{Penny2024}.

Adam Brown and Andrew Fagg's art installation \emph{Bion} uses similar techniques to Penny's but over a larger scale. One thousand individual ``bions'' communicate using IR, making sounds in response to each other and to human presence \cite{Cuzziol2018}. Jon McCormack's \emph{Eden} simulated a virtual ecosystem of sonic agents that, over time, learn to make ``interesting'' sounds that keep the work's human audience engaged \cite{McCormack2007b}. The agents also used sound to communicate with each other, and an evolutionary learning process encouraged multiple successful strategies for how agents should behave in response to different sonic cues. Rafael Lozano-Hemmer's \emph{Linear Atmosphonia} is an interactive sound and light installation comprised of 3000 individual small speakers, each with their own LED lights and distance sensor. Each speaker hangs from the ceiling and detects the presence and movement of people under it.  The scale of the work creates an immense sense of presence and sonic immersion, the sound of ``wind, water, fire, ice, over 200 types of insects, over 300 types of birds, bells, metronomes, bombs'' channelled to hundreds of speakers that float above participant's heads \cite{Hemmer2019}.

\subsection{Soundscape Ecology}
\label{ss:soundscape-ecology}

The term ``soundscape ecology'' was introduced by Pijanowski and colleagues in 2011 \cite{Pijanowski2011}. The authors draw a distinction between prior ecological relationships with sound, such as acoustic ecology \cite{Schafer1977, Truax2001}, which is focused primarily on human-centred relationships between sound and environment.

In such research, the technologies and methods are all focused on passive \emph{listening} to the sonic environment, i.e.~the devices or listeners just \emph{listen,} they don't in general contribute sound to the environment or respond to the calls of animals for communication or other purposes. In contrast, the work introduced in this paper deliberately seeks to intervene in the acoustic environment (both human and animal), populating the sound-space with new sounds created autonomously via machine listening. 

Krause categorised the complex collections of sound in the environment into three distinct areas: biological (``biophony'') -- the sounds made by organisms, ``geophony'' -- natural environmental sounds such as wind, rain and thunder, and lastly ``anthrophony'' -- those caused by humans \cite{Krause1987, Pijanowski2011}. Based on the work described in this paper, we propose the term ``cyberphony'', representing those sounds emanating autonomously from machines. We stress the term ``autonomously'' in this differentiation, indicating the machine has some internal decision-making mechanism that determines what and where sound will be produced. So our categorisation would include devices like smartphones, robots, and \emph{Holon}, but exclude things like car horns\footnote{An exception might be that of an autonomous vehicle that decides to make noise without direct human control.}, emergency vehicle sirens or construction noise. A contemporary urban environment is likely to contain a mix of all four of these categories.

Having given an overview of the conceptual and research background on the uses of sound in human-computer interaction, we now turn to presenting the artwork itself.


\section{Holon}
\label{s:holon}

\begin{quote}
    ``Fertile vistas may open out when commonplace facts are examined from a fresh point of view''
    \flushright --- L.L.~White, quoted in \cite[p.45]{Koestler1967}
\end{quote}

\emph{Holon} is an interactive, solar powered installation of 130 cybernetic organisms that autonomously listen to their environment and communicate using sound. The work overlays the existing soundscape ecology with an introduced species (the cybernetic organism -- an individual \emph{holon}), which amplifies and repeats sounds that might be missed or go unheard by, for example, the casual human observer. 

\begin{figure}[h]
  \centering
  \begin{tabular}{ccc}
       \includegraphics[width=0.3\linewidth]{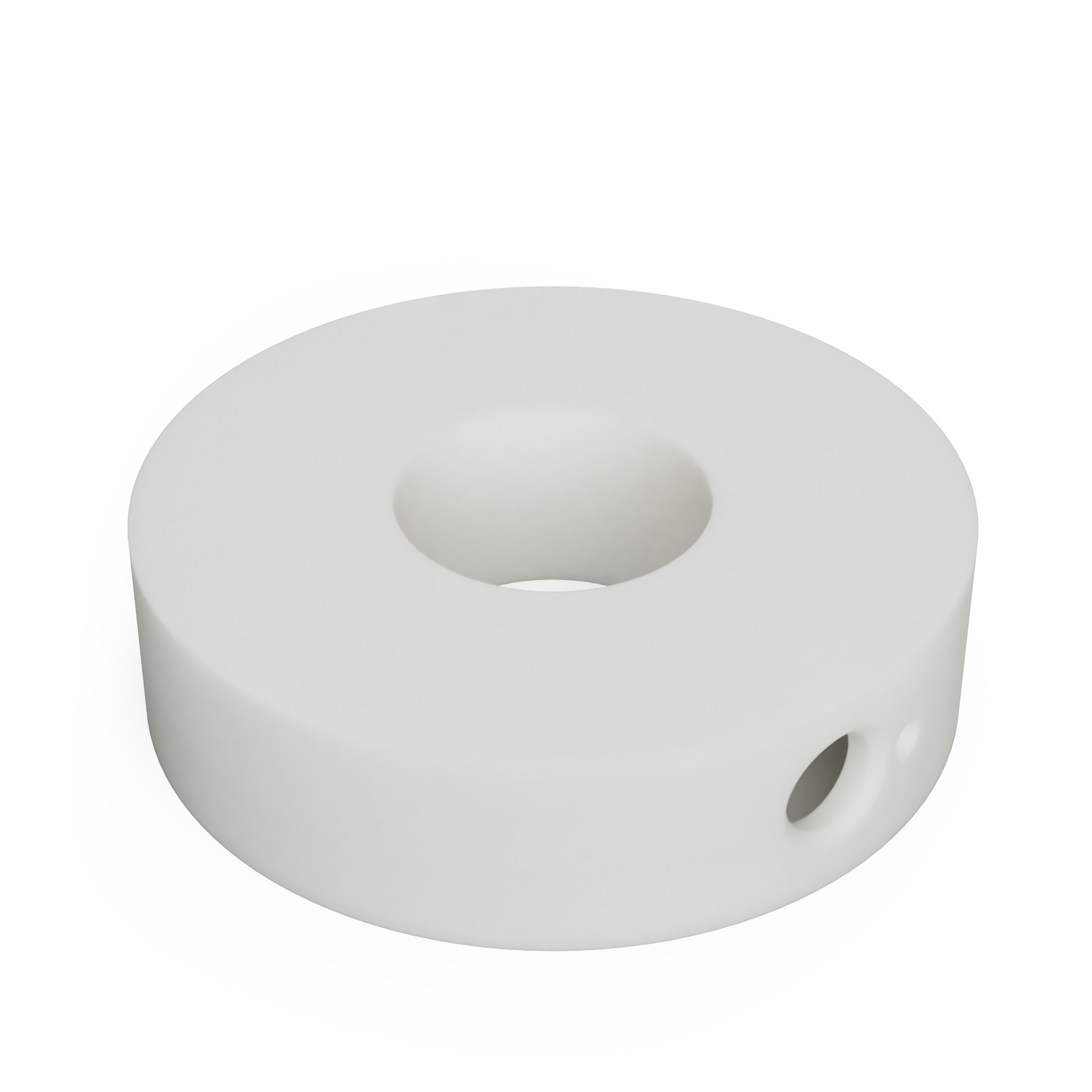} & 
       \includegraphics[width=0.3\linewidth]{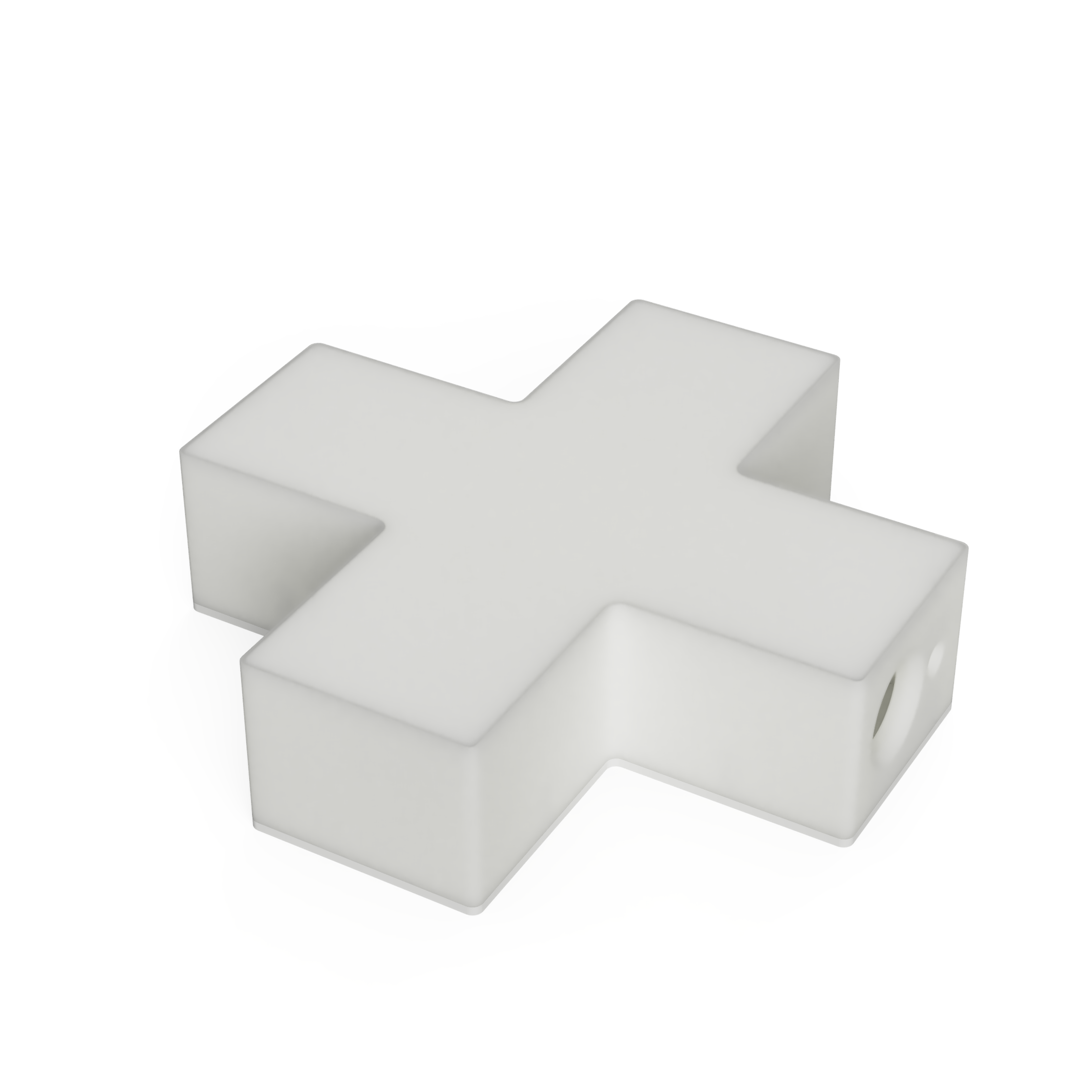} &
       \includegraphics[width=0.3\linewidth]{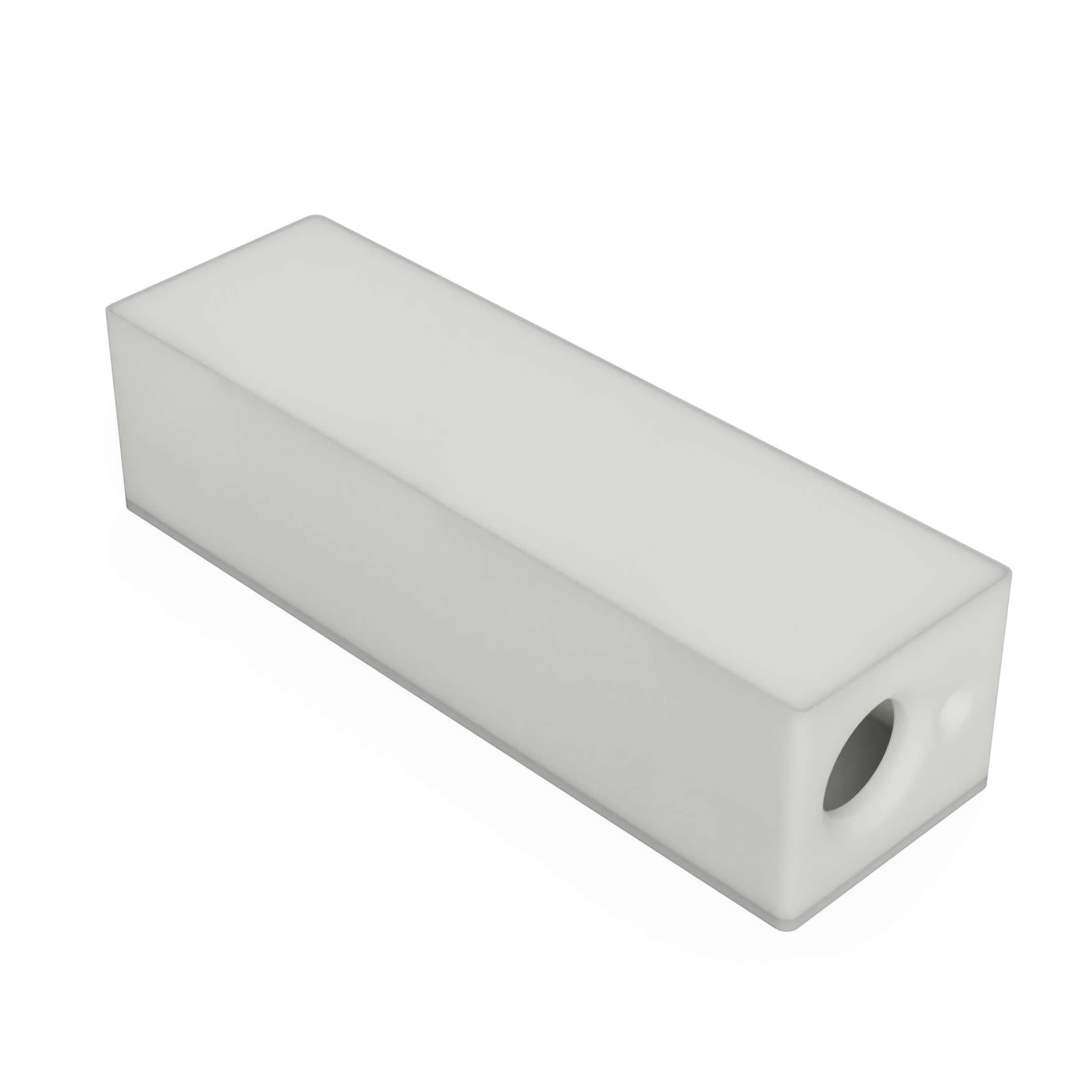} \\
  \end{tabular}
  \caption{The three different holons: composer/generator (left), collector/critic (middle) and disruptor (right). }
  \label{fig:holon-types}
\end{figure}

There are three distinct kinds of holon (Fig.~\ref{fig:holon-types}): the \emph{composer/generator} ($\bigcirc$ shape), the \emph{collector/critic} ($\bigtimes$ shape) and the \emph{disruptors} ($\: \slash \; \text{and} \; \backslash \;$ shapes) -- explained in detail in the next section. Each holon (approximately 35cm $\times$ 35cm $\times$ 6cm) is a self-contained, solar powered physical object attached to a 100cm long mounting pole, allowing it to be strapped to heritage listed pier pylons at an urban waterfront dock (Docklands in Melbourne, Australia). The collection of holons float above the water, signalling and communicating to each other and to other biological species in their vicinity using sound and light (Fig.~\ref{fig:holon}). The shapes reference old maritime navigation signals that were frequently used in the previous century.  

Conceptually, the work is a provocation on the encroaching colonisation of our landscape and soundscape by technology. Each cybernetic creature seeks to communicate with its neighbours via sound. Creatures listen to their acoustic environment and attempt to find sound frequencies that are currently unoccupied by human, machine or animal sounds. Based on the organisational principles outlined by Arthur Koestler in his 1967 book ``The Ghost in the Machine'' \cite{Koestler1967}, each individual holon is part of a collective whole: an network of multiple agencies that includes the biological, geological, computational and human. By ``occupying'' sonic space that may have been previously used by other species, the work reminds us, through a machine-augmented soundscape, of those fauna potentially missing or lost through displacement by urban development or loss due to climate change \cite{Austhreat2023}. Additionally, by listening, playing and modifying sounds from the environment, the work also seeks to augment communication with the non-human species that currently inhabit the work's environment.

\subsection{Technical Details}
\label{ss:holon_technical_details}

\begin{figure}[h]
  \centering
  \includegraphics[width=0.9\linewidth]{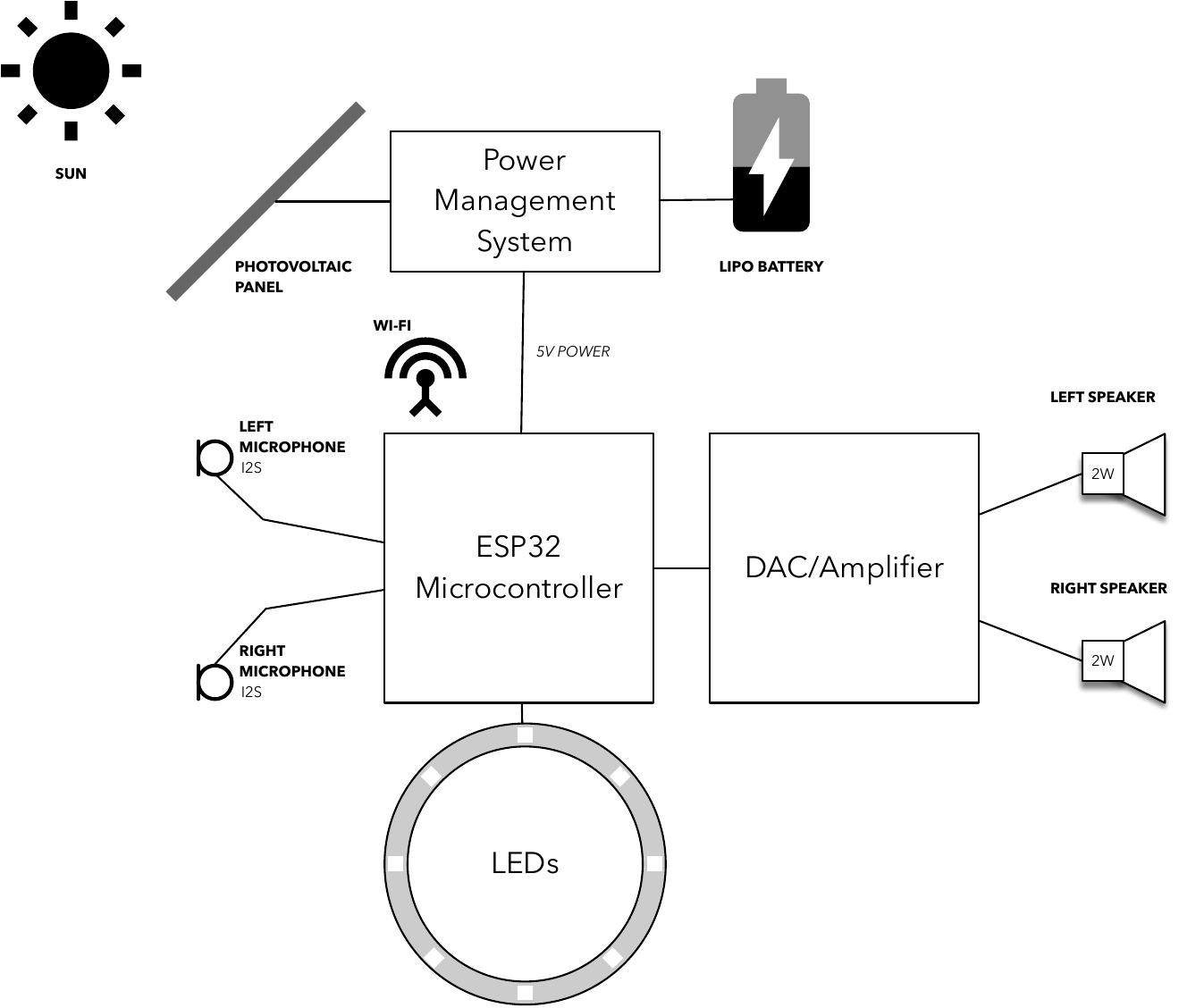}
  \caption{Schematic diagram of an individual holon.}
  \label{fig:holon-schematic}
\end{figure}

\begin{figure*}[h]
  \centering
  \includegraphics[width=\linewidth]{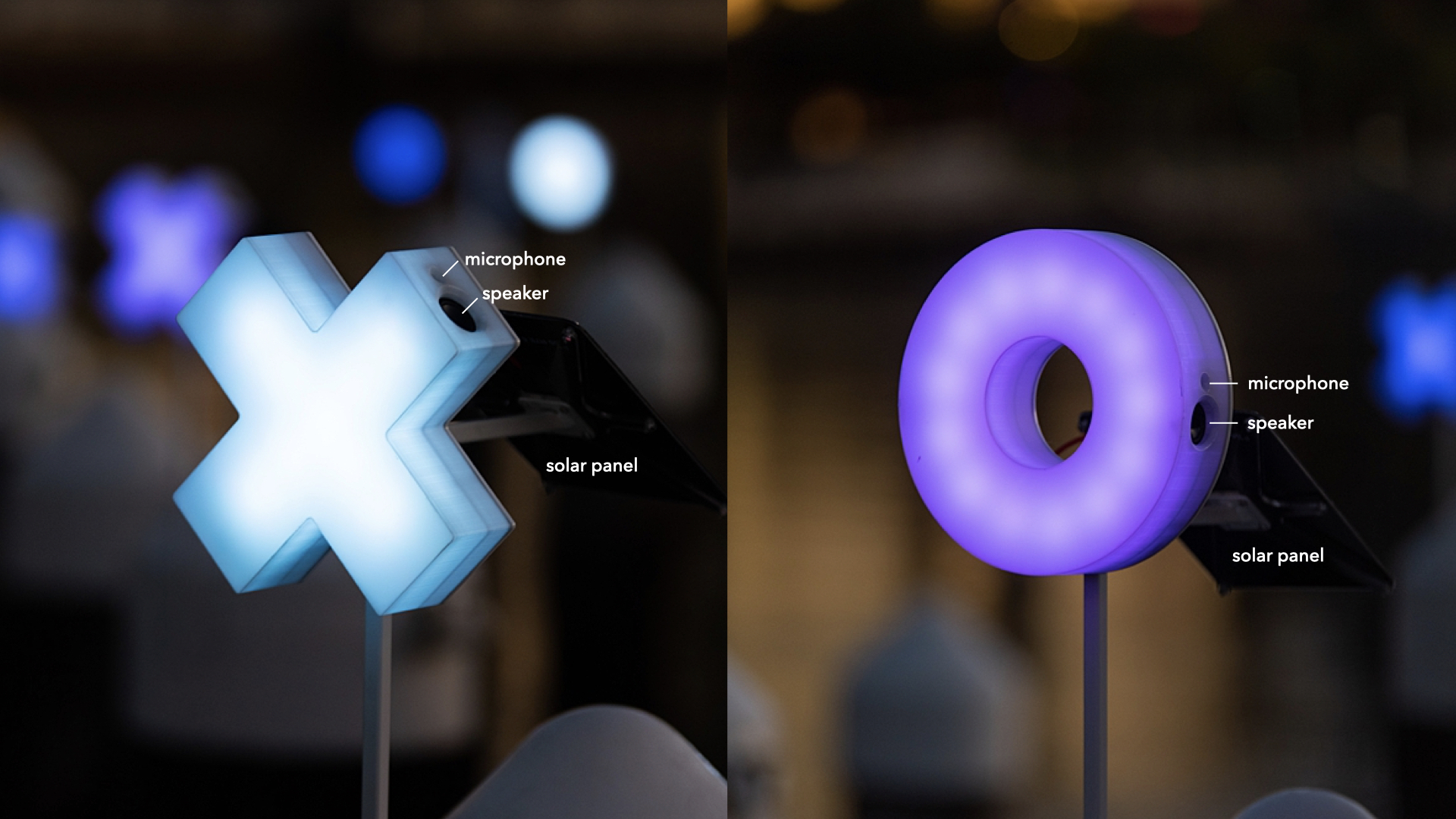}
  \caption{Closeup images of the collector/critic holon (left) and generator/composer holon (right), showing the location of the speaker, microphone and solar panel. Identical microphone/speaker combinations are also present on the opposite side of each (not visible in the figure). }
  \label{fig:holon-diagram}
\end{figure*}

An individual holon is a self-contained cybernetic organism, equipped with multiple microphones, speakers, LED lights and an on-board microcontroller (Fig. \ref{fig:holon-schematic}). Holons get their energy from the sun via a photovoltaic cell attached on their backs (Fig.~\ref{fig:holon-diagram}). Sun charges an internal battery, allowing them to make sound and light after the sun goes down. The more energy they receive during the day, the more lively they become at night.

As introduced in the previous section, there are three types of holon, the type determining their overall behaviour in the soundscape. 
The generator/composer ($\bigcirc$) generates sounds based on its local acoustic environment. Each composer seeks to generate sound in unoccupied frequencies, attempting to claim those frequencies not used by others (here ``others'' refers to the biophony, geophony, anthrophony and cyberphony of the soundscape).
The collector/critic ($\bigtimes$) collects sounds from the environment, passing on those it thinks are novel. Collectors harvest sound and keep samples of sounds they find interesting. Later in the day they emit those sounds in response to the hearing the generator/composers.
The disruptors ($\: \slash \; \& \; \backslash \;$) listen for sounds made by other holons and try to change and disrupt them by modifying the sound before repeating it. Disruptors use digital signal processing (DSP) to modify the sounds they hear.

This model was inspired by previous research on the evolution of birdsong by sexual selection \cite{Catchpole1987, Miller2000a}. Male birds compete for female attention through song, the quality of the song being a proxy for genetic quality. Female birds judge the ``quality'' of the song as a basis for mate selection. The theory (originally proposed by Charles Darwin \cite{Darwin1871}) hypothesised that, over time, the quality of both song and critic evolve as the males compete to become better singers and the females become more nuanced judges\footnote{Since Darwin's time, much evidence has been observed to validate the hypothesis \cite{Miller2000a}.}. Todd and Werner devised a co-evolutionary music composition system of ``composer'' and ``critic'' agents that had the ability to generate complex compositions through co-evolution \cite{Todd1998}.

As discussed in Section \ref{ss:soundscape-ecology}, research pioneered by Krause in soundscape ecology demonstrated that different species tend to fill different regions of the frequency spectrum \cite{Krause1987}, the theory being to avoid conflict between different species using sound through occupation of similar frequencies. Of course sound is also temporal, so multiple species may occupy a similar frequency spectrum, provided that their use is not coincident or unlikely to occur at the same time as other species (a common differential is nocturnal/diurnal behaviour, for example). Research into soundscape ecology has also demonstrated that frequency diversity is a good bio-marker for ecosystem diversity and health \cite{Pijanowski2011, Krause2013}.

The \textbf{composer/generator} holon uses its microphones\footnote{The microphones have a usable frequency range of 20Hz - 16kHz).} to listen to sound, continuously capturing audio frames which are first highpass filtered to retain only frequencies above 80Hz. This filtering is to eliminate low-frequency rumble and certain types of wind noise. Next, an FFT analysis on the incoming sound frames is performed to convert samples from the time domain to the frequency domain. The energy in each frequency bin\footnote{We use the Mel scale \cite{Stevens1937} frequency divisions over 128 bins.} is computed and the long-term running average and range is calculated for each frequency. Over time this information should represent the sonic ``signature'' of the holon's acoustic environment. At night, the holon selects frequencies with the lowest average energy and generates sound at those frequencies, to ``fill the void'' not occupied by sounds at competing frequencies. Sounds are generated from basic sinusoidal waves. The attack and decay of the sound is calculated using a heuristic based on the range of energy heard in the specific band: a high dynamic range results in short attacks and decays (``staccato'' in musical terms), lower dynamic range results in longer attacks and decays (``legarto'').

The \textbf{collector/critic} holon uses its microphones to collect sounds it finds ``interesting'', retaining a collection of the top $n$ most interesting sounds it hears (in exhibition, $n = 32$). Sound samples are streamed in from the microphone, and high-pass filtered at 80Hz as for the composer/generator type holon. An onset detection algorithm is used to separate background noise from audio ``soundmarks'' \cite{Schafer1977} and upon onset detection the holon records the sound into a temporary buffer in on-board memory. Recording continues until the end of the sound is detected (by volume differential), or the record time exceeds 30 seconds. Incoming recordings are then analysed to determine if they should be retained. Analysis is performed using the following measurements:

\begin{enumerate}
    \item The dynamic range (in decibels, dB) of the sample is computed;
    \item The zero-crossings for the sample are computed;
    \item The Mel Frequency Cepstral Coefficients (MFCCs) for the sample are computed;
\end{enumerate}

Together these measurements form the \emph{analysis vector}. To see if a recorded sample warrants keeping (i.e.~is considered ``interesting''), the analysis vector is compared with the list of existing samples previously collected (the holon begins with an empty list, meaning the first recording will always be initially collected). Basic statistical measurements of the list are performed, computing the mean and standard deviation for each analysis measurement. A new sample will be added to the list if (and only if) it increases the standard deviation of the list (i.e.~increases the variety of sounds). If the list is full (list length is determined by available memory, the ESP32-S3 has 8Mb of PSRAM) then the sample with the closest euclidean distance to the incoming sample's analysis vector is marked for replacement.

While more advanced differentiation and identification would be possible using deep learning neural networks for example (e.g. \cite{kahl2021birdnet}), the processing requirements are beyond the low-power capabilities of the microcontroller used, and would significantly increase the power requirements for each holon -- an important consideration when the work is 100\% solar powered. Nonetheless, the relatively simple mechanism described above allows the holon to find a wide variety of different sounds (we observed capturing of tram bells, car horns, bird sounds, rain, thunder claps, human conversation, music, construction sounds, boat noises, waves lapping, insect sounds and the sounds of generator/composer holons during the two week installation period).

The \textbf{disruptor} agents are used to increase novelty into the soundscape. Like the collector/critic holons they listen to the sounds of the environment and use onset detection to register the arrival of a ``soundmark''. Upon detection they transform that sound using DSP techniques to disrupt or inject change to the soundscape. We used a number of basic DSP transformations, including pitch shifting (shifting the frequencies up or down one octave), frequency modulation using a random fixed frequency, and ring modulation (where a carrier wave is combined with a modular waveform). The overall effect is to transform the soundscape and prevent the holon collective from settling into a stasis. We found that the disruptor agents helped shift the composer/generator agents from constantly playing similar frequencies.

\subsection{Operation and Results}
\label{ssoperation-results}

\begin{figure}[h]
  \centering
  \begin{tabular}{cc}
       \includegraphics[width=0.25\linewidth]{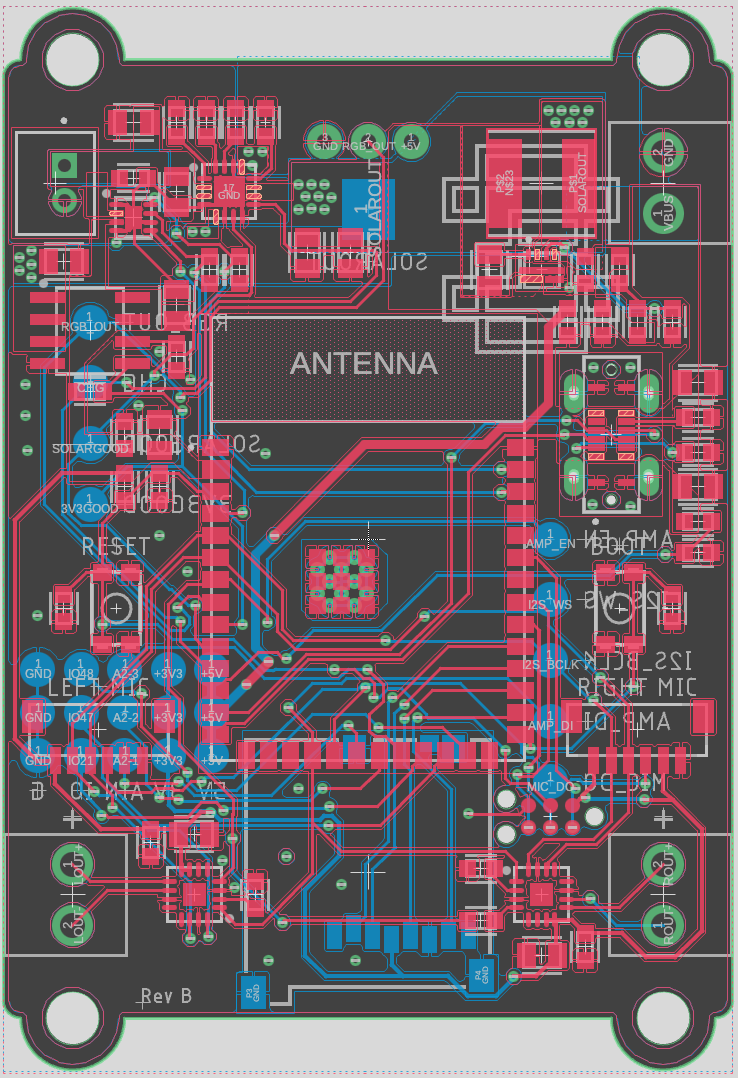} & 
       \includegraphics[width=0.4\linewidth]{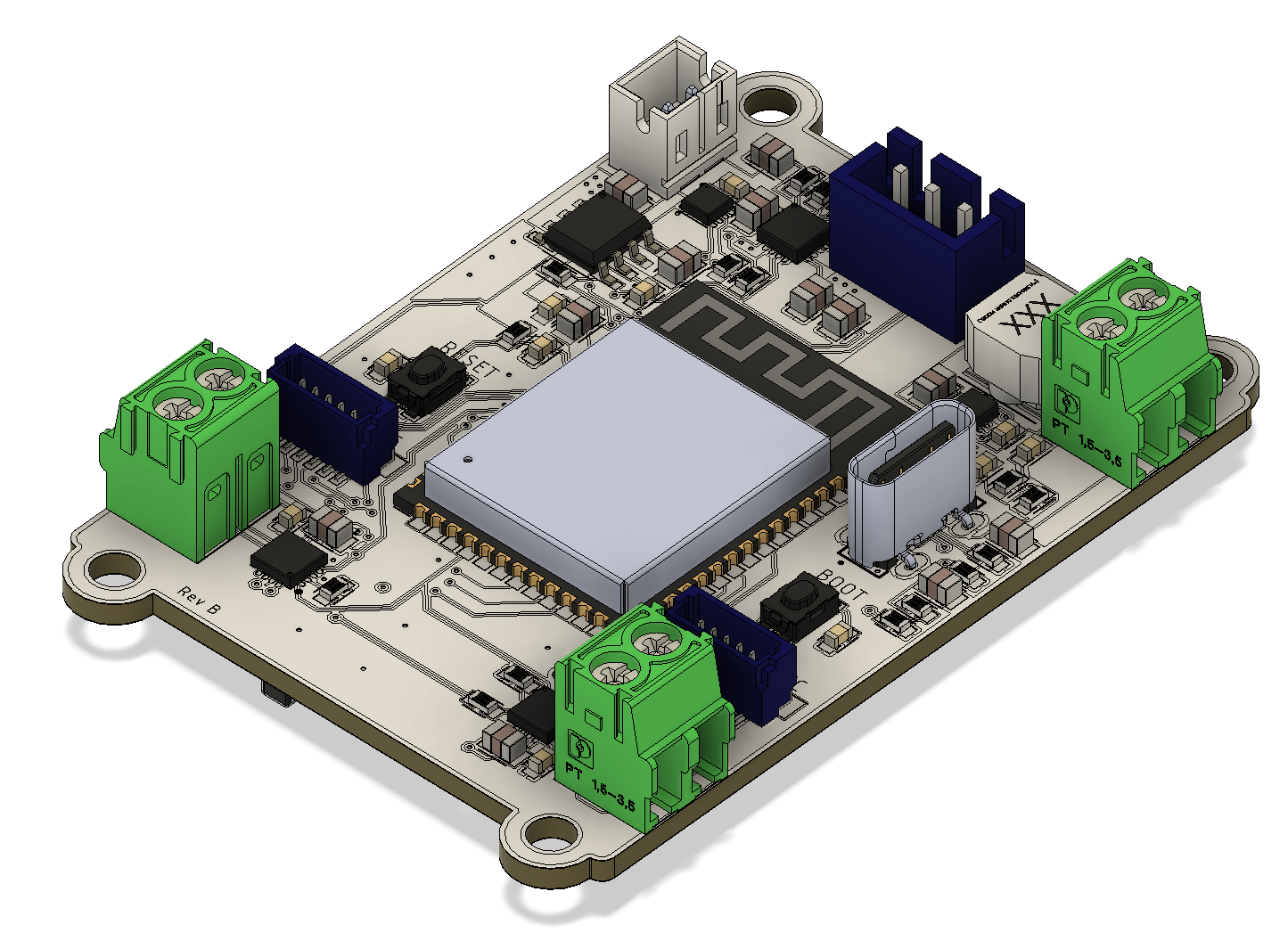} \\
  \end{tabular}
  \caption{The Holon circuit broad design (left) and assembled PCB (CAD rendering, right) }
  \label{fig:holon-electronics}
\end{figure}

We designed and built 130 individual holons (50 composer/generators, 50 collector/critics and 30 disruptors) in our university lab. The physical form was fabricated using a combination of 3D printing and laser cut acrylic. Care was taken to ensure the devices were water-sealed, using rubber sealing between parts and hydrophobic material to prevent water ingress into the microphone and speaker outlets. We also designed the circuit board and electronics, which included the solar power management system, microcontroler and audio amplifier. Each holon was assembled by hand and tested before being deployed on site.

The installation was installed on a waterfront dock for 19 days as part of a city-funded public art festival (Now or Never Arts Festival). Over 70,000 people visited the installation during this time. Visitors generally expressed a wonder and delight at seeing and hearing such a strange and unusual intervention, with many commenting on how the work collectively gave the impression of being ``alive''. Regular visitors (such as local residents and nearby office workers) commented on the different behaviours at different times of the day: during the day the composer/generators would often ``sing'' together, generating an ambient soundscape that could be heard around the general area of the work. People often encountered this soundscape before sighting the physical installation itself. At night the collector/critics and disruptors become more active, changing the soundscape to reflect the increased activity. Each holon also had built-in LED lights that reflected the their current state. They would light up when making a sound, so as to direct any observer's attention to the source of their sound. The collector/critic holons become red when acquiring a sound and would glow blue for a short time after if they accepted this new sound as ``interesting''. Disruptors would change intensity when disrupting in time with the modulation being applied to the audio signal output.

As part of the exhibition we collected data on what each Holon was doing over the course of the exhibition period (19 days), using information sent on a closed WiFi network as a means of communicating with each device remotely for the purposes of monitoring behaviour. It is important to stress that the only communication between holon's was via the sound they made. The WiFi connectivity was used only for analysis and setting system parameters. The collected data was used to analyse performance over an extended period for the purposes of sustainability research (energy consumption vs energy generation) and behaviour (how well the holons adapted to the sonic environment). As the sustainability performance is outside the scope of this paper we only report the sound behaviour results here.

\begin{figure}[h]
  \centering
       \includegraphics[width=\linewidth]{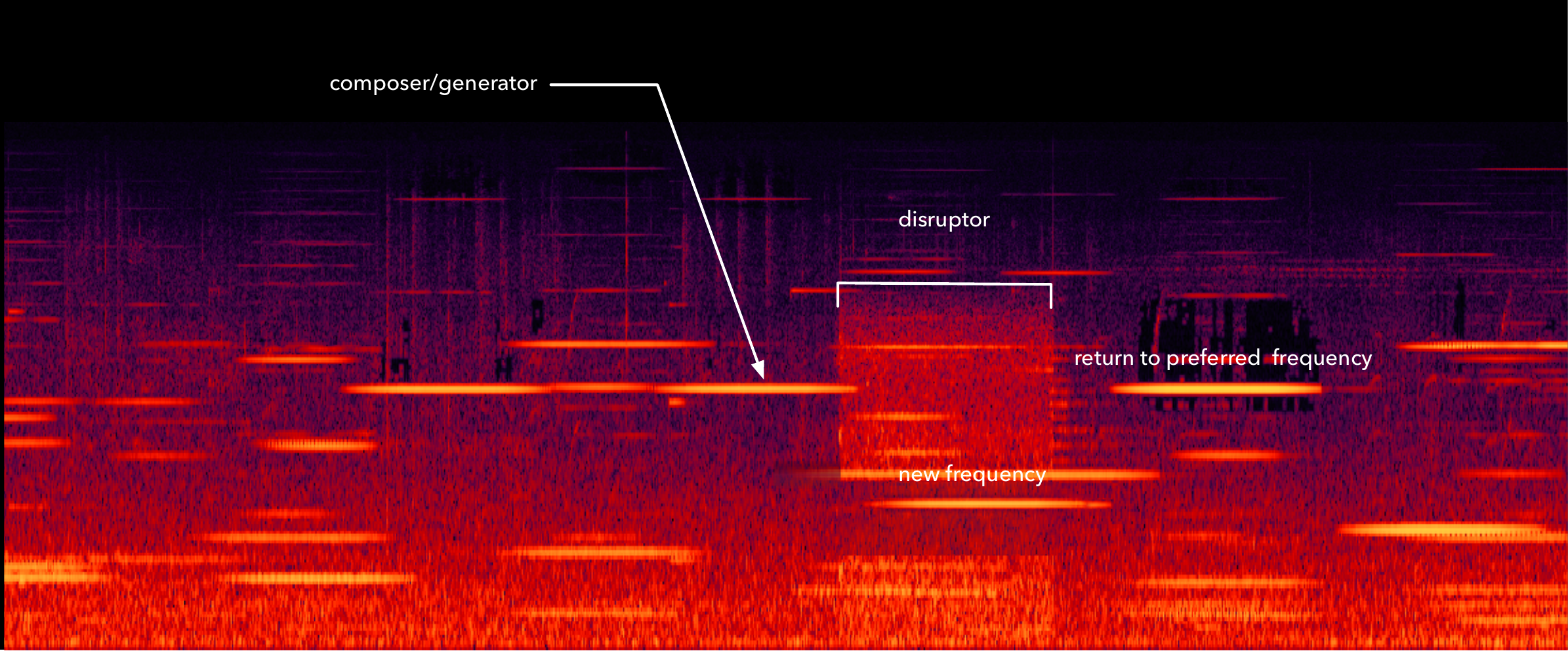}
  \caption{Audio spectragram showing how composer/generator holons occupy specific frequencies. The appearance of a disruptor forces the generator to switch frequencies. The general ``noise'' in the spectragram represents ambient and background noise, such as distant traffic and construction, which tends to be loudest in the lower frequencies.}
  \label{fig:holon-spectra}
\end{figure}

In general, the system was able to discover and make use of frequencies unoccupied in the existing soundscape. Figure \ref{fig:holon-spectra} shows a spectrgram from captured audio from several holons over a period of around 10 minutes. The horizontal orange lines in the figure are indicate individual holons that have found frequencies unoccupied by other sounds (including those of other holons). In this particular sequence, a disruptor tries to occupy a wide range of frequencies, with particular intensity around the frequency of the highlighted generator. The generator responds to this signal by temporarily shifting the frequency of its generated sound in order to claim less occupied sonic territory. Once the disruptor stops, the generator returns to its preferred original frequency.

\begin{figure}[h]
  \centering
       \includegraphics[width=\linewidth]{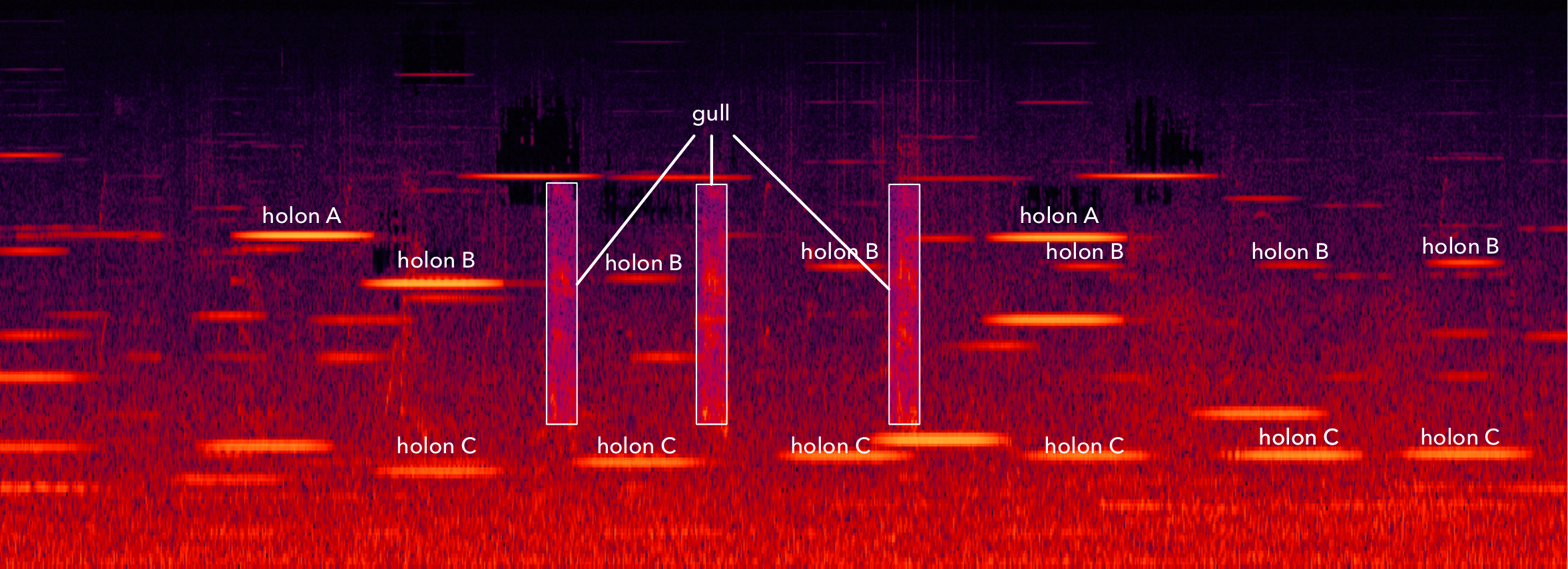}
  \caption{Audio spectragram showing the adaptation of holons to a silver gull.}
  \label{fig:holon-spectra-birds}
\end{figure}

Holon also adapted to non-human sounds within the environment. Fig.~\ref{fig:holon-spectra-birds} shows the spectragram of a sequence 
when a silver gull \emph{(Chroicocephalus novaehollandiae)} was nesting in the area around several holons. The three vertical bars (highlighted in white) show the gull's three cries, occupying a significant range of frequencies. The figure shows the behaviours of different holons: holon A, while within the gull's frequency range is avoiding collisions by waiting for the gull frequencies to stop. Holon B adjusts timing and pitch (moving slowly upward). There appears to be a kind of call and response exchange between holon B and bird -- each choosing a moment when the other is silent to make sound (whether this is co-incidence or not requires further investigation). Lastly, Holon C has found a frequency outside those of the gull so continues to make sound in parallel.

More generally, we observed a variety of bird life that made use of the general area around the work. Over the exhibition period we undertook an audit of the birds nesting or gathering around the work at dusk and in the early evening and found that while the variety of different species observed remained the same, the number of birds increased during the 19 day exhibition period by around 20\%. While not by any means conclusive on the power of the work to attract non-human species, we can at least be confident in claiming that it did not diminish the existing bird populations while attracting a large number of \emph{homo sapiens}.

\section{Conclusion}
\label{s:conclusion}

This paper presented both the technical and conceptual foundations of \emph{Holon}, an artwork which features the novel use of sound as its primary interface between different agents in a complex urban ecosystem.

\begin{figure}[h]
  \centering
       \includegraphics[width=\linewidth]{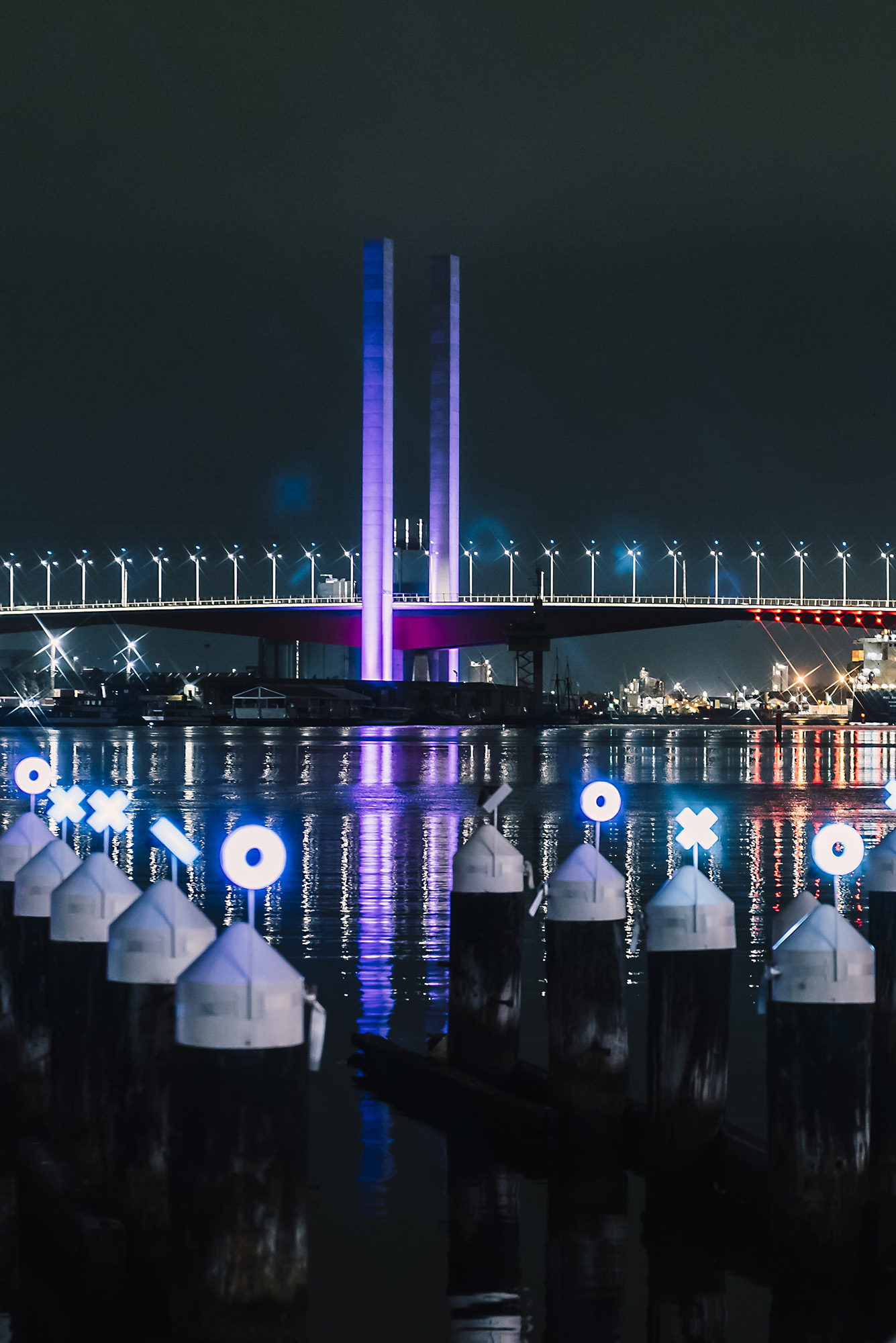}
  \caption{Close up of Holon at at night showing the arrangement of composer/generators, collector/critics and disruptors in situ.}
  \label{fig:holon-arttrail}
\end{figure}

Holon extends the concept of sound-based communication beyond that of human and machine to include other forms of organic life. Holon's ability to generate soundscapes specific to place and acoustic environment makes it unique. Rather than using sound for monitoring, sonification, or composition, the work simultaneously occupies and alters the sonic landscape in which it is placed. The work is an example of sound-based interaction between human, animal and cybernetic agents where each is considered an equal participant. As discussed, this opens interesting new possibilities for communication between non-human life and technology. From a research perspective, it suggests richer and deeper interactions between technological systems and biological life, while also asking the human listener to be more aware of their soundscape ecology: to listen to the voices of things beyond the human.

Holon suggests a possible future where electronic life seeks to co-exist with biological life, both as a replacement for the nature we have displaced or destroyed and the new nature of bio-machine cybernetics.

\section{Acknowledgements}
Holon was commissioned by the \emph{Now or Never Festival}, City of Melbourne in collaboration with Experimenta. The work was designed and build at SensiLab. Thanks to Camilo Cruz Gambardella, Monika Schwarz and Sam Trolland who helped with the building, the team at Experimenta, including Kelli Alred, Ciaren Begley, Stephanie David, Kim De Krester, Jeannie Mueller and Lubi Thomas. Thanks to the City of Melbourne,  Annette Vieusseux Executive Producer of Now or Never.

\bibliographystyle{isea}
\bibliography{isea}

\section{Author(s) Biographies)}
Jon McCormack is an Australian artist and academic. He is the founder and directore of SensiLab, a creative technologies research centre based at Monash University in Melbourne.

Elliott Wilson is an electronics designer, developer and maker. He is SensiLab's Lab Manager.

\end{document}